\documentclass[reprint,superscriptaddress,prl,showpacs]{revtex4-1}
\usepackage{graphicx}
\usepackage[usenames]{color}
\usepackage{multirow}
\usepackage{verbatim}
\usepackage{amsmath}
\usepackage{amsfonts}
\usepackage{amssymb}
\usepackage{mathrsfs}
\usepackage{url}
\usepackage[justification=justified,singlelinecheck=false]{caption}
\usepackage{subcaption}
\usepackage{float}

\begin{document}

\title{Adsorption-Induced Solvent-Based Electrostatic Gating of Charge Transport through Molecular Junctions}
\author{Michele~Kotiuga}
\email{mkotiuga@berkeley.edu}
\affiliation{Department of Physics, University of California, Berkeley, CA, USA}
\affiliation{Molecular Foundry, Lawrence Berkeley National Laboratory, Berkeley, CA, USA}

\author{Pierre~Darancet}
\email{pdarancet@lbl.gov}
\affiliation{Molecular Foundry, Lawrence Berkeley National Laboratory, Berkeley, CA, USA}
\affiliation{Department of Applied Physics and Applied Mathematics, Columbia University, New York, NY, USA}

\author{Carlos~R.~Arroyo}
\affiliation{Department of Applied Physics and Applied Mathematics, Columbia University, New York, NY, USA}

\author{Latha~Venkataraman}
\email{lv2117@columbia.edu}
\affiliation{Department of Applied Physics and Applied Mathematics, Columbia University, New York, NY, USA}

\author{Jeffrey~B.~Neaton}
\email{jbneaton@lbl.gov}
\affiliation{Department of Physics, University of California, Berkeley, CA, USA}
\affiliation{Molecular Foundry, Lawrence Berkeley National Laboratory, Berkeley, CA, USA}
\affiliation{Kavli Energy NanoSciences Institute at Berkeley, Berkeley, CA, USA}
\email{jbneaton@lbl.gov}

\begin{abstract}
Recent experiments have shown that transport properties of molecular-scale 
devices can be reversibly altered by the surrounding solvent. Here, we use a 
combination of first-principles calculations and experiment to explain this 
change in transport properties through a shift in the local electrostatic 
potential at the junction caused by nearby conducting and solvent molecules 
chemically bound to the electrodes. This effect is found to alter the 
conductance of 4,4'-bipyridine-gold junctions by more than 50\%. Moreover, 
we develop a general electrostatic model that quantitatively predicts the 
relationship between conductance and the binding energies and dipoles of 
the solvent and conducting molecules. Our work shows that solvent-induced 
effects are a viable route for controlling charge and energy transport at 
molecular-scale interfaces.
\end{abstract}
\maketitle
Single-molecule junctions, individual molecules contacted with macroscopic 
electrodes, provide unique insight into the nanoscale physics of charge, 
spin, and energy transport~\cite{LathaRev,NatelsonRev,NitzanRev,MPRev}. To 
date, the most robust and reproducible approach to assemble single-molecule 
junctions is the scanning tunneling microscope-based break junction (STM-BJ) 
technique~\cite{Tao,VenkataramanNature}, allowing statistically significant 
measurements of molecular junction conductance
~\cite{Tao,VenkataramanNature,
BreakJunctionExperiments1,BreakJunctionExperiments2,BreakJunctionExperiments3}, 
thermopower~\cite{Thermopower1,Thermopower2,ThermopowerWidawsky}, mechanical 
properties~\cite{Frei,TaoMech,BJVdW}, and binding mechanisms~\cite{BJVdW}. 
Previously-developed theoretical approaches have led to quantitative agreement 
with experiment for molecular junctions, given a good approximation to the 
junction geometry and a good estimate of the differences in energy $\Delta E$ 
between the junction Fermi energy, $E_F$ and the orbital energy of the 
frontier orbital, either the highest occupied or lowest unoccupied molecular 
orbital (HOMO or LUMO, respectively)~\cite{Kim}. Theoretical works focusing 
on this level alignment~\cite{SanvitoBDT,Thyg} have led to increased 
understanding and control of molecular junction conductance and thermopower 
in terms of junction level alignment, with significant impact on experiments
~\cite{Thermopower1, Thermopower2, Malen, ThermopowerWidawsky}. 

Commonly, these experiments take place at room temperature in a non-conductive 
solvent
~\cite{Tao,VenkataramanNature,BreakJunctionExperiments1,
BreakJunctionExperiments2}, 
which has recently been shown to influence both 
junction formation probability and, in some cases, to alter the conductance
~\cite{Fatemi}.  Despite its practical importance, the impact of these solvents 
on conductance has not yet been fully understood or explained by theory, in 
part due to the large computation cost~\cite{Higgins}, and, therefore, 
continues to be elusive to control. Previous theoretical works have focused 
on the effect of solvent on the average ~\cite{Baldea2012,Baldea2013} and 
dynamical~\cite{Cummings} molecular junction geometries, and how they affect 
level alignment and modify the conductance~\cite{Sanvito}. Another focus has 
been the coupling of transmission channels due to intermolecular hopping 
between conducting molecules~\cite{Reuter}, in the case solvent would influence 
the formation of multiple simultaneous junctions. However a detailed physical 
picture and quantitative framework for understanding how solvent affects junction 
level alignment remains elusive despite clear evidence~\cite{Fatemi}: new theory 
and models are required to understand and better control solvent effects on 
junction transport properties. Given the significant recent interest in 
solvent-base gating of correlated oxides~\cite{Oxide}, graphene~\cite{Graphene}, 
and transition metal dichacogenides~\cite{Chal}, such a theory will have 
general implications.
\begin{figure*}[ht!]
\flushleft
\includegraphics[width=\textwidth]{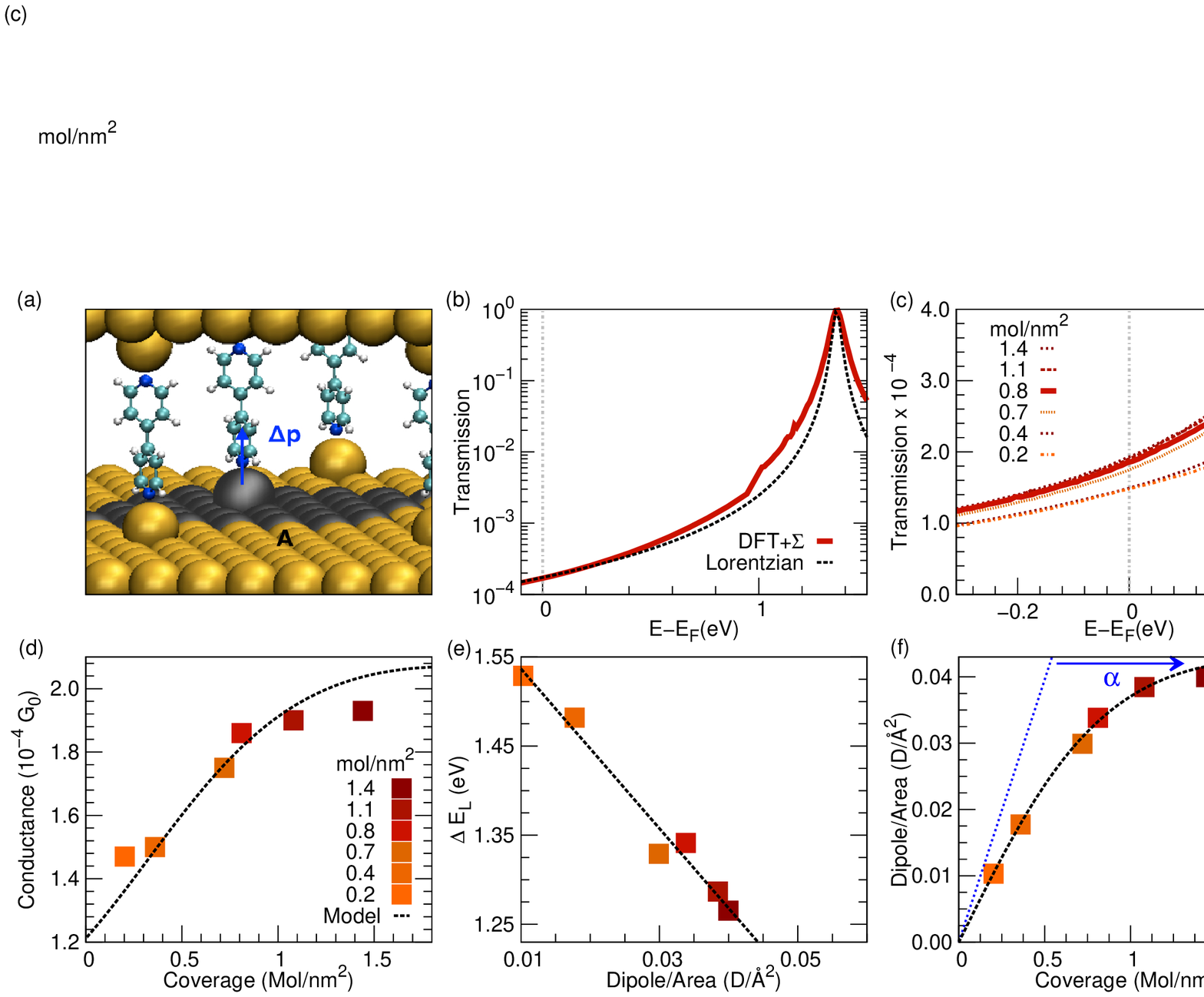} 
\caption{(a)  Junction geometry for 0.8 molecule/nm$^2$ coverage with periodic 
boundary conditions used in transmission calculations. Cross-sectional area 
indicated with a black shadow, and the induced dipole with a blue arrow. (b) 
Transmission function of Au-BP-Au junction at a 0.8 molecule/nm$^2$ coverage 
with Lorentzian fit. (c) The transmission function of Au-BP-Au junctions for 
different coverages ranging 0.2-1.4 molecule/nm$^2$ near $E_F$. (d)  The DFT+
$\Sigma$ conductance of Au-BP-Au compared to a electrostatic based model as 
a function of coverage.  (e) The linear dependence of level alignment on 
dipole/area. (f) The nonlinear dependence of dipole/area on coverage: this 
is due to depolarization of the polarizable interface dipoles, black line 
from model, blue is the case without depolarization.}
\label{fig1}
\end{figure*}

In this Letter, we explain the effect of solvent on molecular device transport 
properties in a manner analogous to a chemical electrostatic gate controllably 
altering the local potential of the junction. We demonstrate how the electrode 
surface can act as a template to order the adsorbate molecules near the 
junction, resulting in large, ordered induced dipoles and a sizable, coherent 
shift of the average junction electrostatic potential, outweighing bulk 
effects associated with thermal fluctuations at room temperature and the low 
intrinsic dipole moment of the unbound solvent molecules. This picture arises 
out of our explicit quantitative theoretical calculations of the effects of 
the molecular coverage, $\theta$, on the transport properties of 4,4’ 
bipyridine-gold (Au-BP-Au) junctions comparing directly with new 
experimental data. For Au-BP-Au junctions, we accomplish this by taking 
advantage of the periodic boundary conditions in our transport calculation by 
changing the cross-section of the supercell - see Figure  \ref{fig1}(a) - 
and find that a high molecular coverage results in conductance values per 
molecule one and a half times larger than those in the dilute limit. This 
effect originates from the coverage dependence of the local potential at 
the junction~\cite{Selloni}, which acts as a local electrostatic gate and 
alters junction level alignment. We further introduce a model that, in 
combination with first-principles calculations, explains the experiments 
quantitatively for arbitrary surface concentrations of solvent and 
conducting molecules and predicts that the conductance of HOMO- and LUMO-
conducting molecular junctions exhibits an opposite trend in solvent-
dependence, demonstrating the potential for different solvents to 
discriminate between hole or electron transport, much like a thermopower 
measurement. The magnitude of these effects is comparable to the one 
induced by ionic gating~\cite{Capozzi}, establishing liquid neutral 
solvents as a potential route to realize three-terminal device physics 
in single-molecular junctions.

We optimize junction geometries using density functional theory (DFT) within 
the generalized gradient approximation (GGA) of Perdew, Burke, and Ernzerhof 
(PBE)~\cite{PBE1,PBE2} using the SIESTA code~\cite{Siesta1,Siesta2}. Each 
gold lead is modeled with six (111) layers, where the three outer layers of 
each are constrained to the bulk geometry and the cross-section of gold 
atoms in the lead supercell ranges from 3x3 to 8x8. The Au-BP-Au junction 
geometry used in the caluclation, as shown in Figure \ref{fig1}(a), is 
similar to previous work~\cite{SuYingNatNano}. Our transport calculations 
are carried out within an {\em ab initio} Landauer framework using a 
scattering state method implemented in the Scarlet code~\cite{Scarlet}, with 
an in-plane $k$-mesh of 16x16 for the 4x4 cross-section and is adjusted 
accordingly for other junctions. In order to correct the Kohn-Sham DFT-PBE 
energy level alignment for missing exchange and correlation effects, we 
apply an approximate GW correction within the DFT+$\Sigma$ framework~\cite
{SuYingNLetters,NeatonPRL} (All details of calculations given in Supp. Mat.).  
We find the Au-BP-Au junction LUMO resonance energy by diagonalizing the 
junction Hamiltonian projected on the molecular subspace and the dipole per 
area (areal dipole) is calculated by integrating  the  first moment of the 
DFT-PBE electronic density on one side of the center of charge of the 
junction,  making use of its near-inversion symmetry. To isolate the induced 
areal dipole upon binding, $\Delta p$, the areal dipoles of the isolated 
gold electrode and molecule are subtracted from that of the junction; see 
Figure \ref{fig1}(a).

The results of our calculations are summarized in Figure \ref{fig1}. Figure 
\ref{fig1}(b) shows the DFT+$\Sigma$ transmission function, $T(E)$, near the 
$E_F$, which is dominated by single LUMO resonance and has a Lorentzian 
lineshape: $T(E)=\Gamma^2/(\Gamma^2+ 4(\Delta E_L-E)^2)$, where $\Gamma$ is 
the full width half maximum ($\Gamma$ = 0.036 eV, see Supp. Mat.) and 
$\Delta E_L\sim 1.3$ eV is the LUMO resonance peak energy relative to $E_F$. 
This agrees with previous experimental and theoretical works on the Lorentzian 
lineshape of the transmission function of Au-BP-Au junctions
~\cite{BPlor,SuYingNatNano,Bagrets,Kim,ThermopowerWidawsky}. Upon variation 
of the molecular coverage over from 0.2 to 1.4 molecule/nm$^2$, the predicted 
conductance ranges from  $1.47-1.93\times 10^{-4} G_0$, shown in Figure 
\ref{fig1}(c), where the conductance $G = G_0T(E_F)$ and $G_0=2e^2/h$, and 
furthermore is shown to vary nonlinearly - see Figure~\ref{fig1}(d). This 
significant variation is not a consequence of intermolecular coupling
~\cite{Reuter}, as at full molecular coverage the in-plane dispersion of the 
LUMO in momentum space is  0.007 eV, accounting for less than a 1\% spread 
in conductance. Instead, the coverage alters conductance much in the manner 
adsorbates on a surface alter the work function~\cite{Kronik}.

As we will now demonstrate, the variation in conductance comes from a shift 
of the local potential. The change in local potential upon coverage can be 
observed in the variation of the LUMO resonance energy as shown in Figure 
\ref{fig1}(e). As BP and the solvent in this work are neutral species, the 
local potential varies at first order as the function of the areal dipole 
$\Delta p/A$ of the system. As shown in Figure \ref{fig1}(e), this 
truncation of the local potential at the dipole term is in excellent 
agreement with the DFT+$\Sigma$ calculations, \textit{i.e} the LUMO 
resonance energy, $\Delta E_L$, as a function of coverage is: 
\begin{align}
\Delta E_L\left(\frac{\Delta p}{A}\right) = \Delta E_L^{\theta \rightarrow 0}-
\eta\frac{\Delta p}{A}.
\label{E0lin}
\end{align} 

The position of the LUMO resonance in the infinitely dilute limit is 
$\Delta E_L^{\theta \rightarrow 0} = 1.63$ eV and the slope is $\eta = 8.95$ 
eV$\cdot$\AA$^2$/D. The slope of this line is less than one expects from 
Poisson's equation, ~$\sim$12$\pi$ in these units (See Supp. Mat.), as the 
two sides of the junction are close enough to interact with each other 
effectively allowing the gold leads to heavily screen the electrostatic 
interactions in the junction. We note that the areal dipole does not vary 
linearly  with molecular coverage due to significant depolarization effects
~\cite{Kronik,macdonald}, as shown in Figure \ref{fig1}(f).

\begin{figure}[t!]
\includegraphics[width=.35\textwidth]{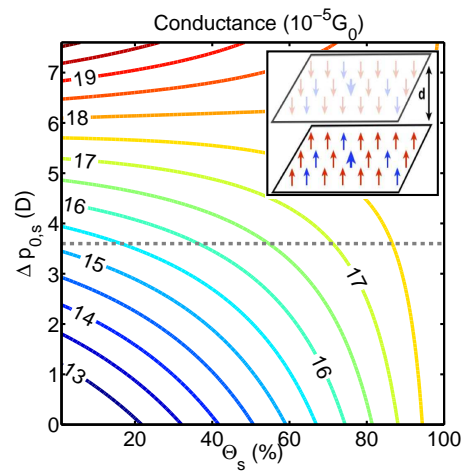}

\caption{ The conductance, given by the electrostatic model, as a function 
of the ratio of BP molecules to solvent on surface concentration, 
$\Theta_s$ and dipole of bound solvent with a molecular coverage of 0.8 
molecule/nm$^2$. Grey dashed line denotes dipole of tricholorbenzene (TCB). 
Inset: Schematic of dipole lattice model with 30\% surface concentration of 
conducting molecule (blue) in solvent (red). The larger arrows denote the 
molecule bridging the interfaces.}
\label{fig2}
\end{figure}

To gain further quantitative insight into the dependence of $\Delta p$ on 
coverage, we develop a lattice model of point-polarizable dipoles induced 
by binding events for both slab and junction geometries. Assuming flat 
electrodes in the vicinity of the junction, the model is comprised of two 
infinite planar interfaces on either side of the conducting molecule of 
length $d$, one at $z=0$ and the other at $z=d$. Each interface has an 
array of dipoles, $\Delta p$, pointing towards the center of the junction 
on a regular rectangular lattice, with spacing corresponding to a given 
molecular coverage; see the inset of Figure \ref{fig2}. The interface 
dipole induced by molecular binding (on either side) can be considered as 
a bare dipole, $\Delta p_0$, with polarizability, $\alpha$, which 
depolarizes due to the collective electric field of all other dipoles, 
and which we express as
\begin{align}
\Delta p(\theta,d)=\frac{\Delta p_0}{1+\alpha k(\theta,d)\theta^{3/2}}.
\label{dep}
\end{align}
where $k(\theta,d)$ is a coverage-dependent sum over purely geometric 
factors (see Supp. Mat.),  following Topping~\cite{Topping}. To fit $\alpha$ 
and $\Delta p_{0}$ we use Eq. (\ref{dep}) in the one-interface limit: 
$d\rightarrow\infty$ and the inducted areal dipoles from DFT calculations 
of BP-Au slabs at various $\theta$. Subsequently, $d$ is fit by using Eq. 
(\ref{dep}) in the case of finite $d$ and induced areal dipoles from the 
corresponding DFT calculations of Au-BP-Au junctions. For the case of BP 
we find $\alpha_{BP}$ = 97 \AA$^3$, $\Delta p_{0,BP}$ = 7.6 D and $d$ = 
7.31 \AA. Combining Eqs. (\ref{E0lin}) and (\ref{dep}), we complete our 
description of how level alignment varies with coverage. Assuming a 
Lorentzian limit for the conductance, which is valid for Au-BP-Au junctions, 
we can use this description of level alignment as a function of coverage 
to and model conductance as a function of coverage as
\begin{align}
G(\theta) = \frac{2e^2}{h}\frac{\Gamma^2}{\Gamma^2+4\left(\Delta 
E^{\theta\rightarrow 0}_L-\eta\theta\frac{\Delta p_0}{1+\alpha k\theta^{3/2}}
\right)^2}.
\label{Gtheta}
\end{align}
This model shows excellent agreement to the first-principles calculations 
in Figure~\ref{fig1}(d) for Au-BP-Au junctions.

To incorporate the effects of solvent, we allow for two species of dipoles 
in our model: one representing the bound solvent molecule and the other 
the bound conducting molecule; see inset of Figure \ref{fig2}.  The 
polarizability, $\alpha_{s}$, and bare dipole, $\Delta p_{0,s}$, of the 
bound solvent are calculated, as for the conducting molecule, by fitting 
the  dipole from solvent on gold DFT calculations at various $\theta$ to 
the one interface depolarization expression - Eq. (\ref{dep}). Breaking 
our infinite planar interfaces into large supercells of $N$ sites, we 
construct a random configuration of the two-dipole species with a specified 
surface concentration of BP molecules to solvent molecule: $\Theta_s$. The 
depolarization of the $N$ dipoles can be written as a system of equations; 
{\em i.e},
\begin{align}
\Delta p_{i}=\Delta p_{0,i}-\alpha_i\sum_{j=1}^N \Delta p_{i}k(\theta, d,r_{ij})
\theta^{3/2},
\end{align}
where $i=1,...,N$, and $k(\theta, d,r_{ij})$ is again dependent on geometric 
factors: $\theta$ is the density of available binding sites and $r_{ij}$ is 
the relative displacement between $\Delta p_i$ and  $\Delta p_j$. Upon solving 
this system of equations, we calculate the average areal dipole per supercell, 
$\Delta \tilde{p}(\Theta_s)$, and can generalize Eq. (\ref{Gtheta}) by 
replacing the depolarized dipole as expressed in Eq. (\ref{dep}) with  
$\Delta \tilde{p}(\Theta_s)$.

\begin{figure}[t!]
\includegraphics[width=.45\textwidth]{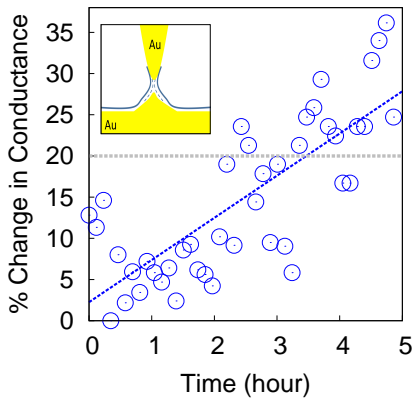}

\caption{Time resolved percentage change in conductance of Au-BP-Au STM BJ in 
TCB over 5 hours. As the solvent evaporates the solution becomes more 
concentrated, and the conductance increases. Comparison to maximal change in 
model shown in grey, inset: cartoon of experiment.}
\label{fig3}
\end{figure}

Figure \ref{fig2} shows the results of our model for conductance as we vary  
$\Delta p_{0,s}$ and $\Theta_s$, using parameter for BP and $\alpha_s$ = 55  
\AA$^3$, assuming and a density of 0.8 available binding sites/nm$^2$. For 
$\Delta p_{0,s} = 0$ we recover our single dipole case with BP at various 
coverages. We note that in the case of high solvent molecule coverage, and 
a large solvent-induced dipole relative to the conducting molecule, we can, 
in principle, use this ``solvent gate'' to achieve a higher conductance 
than for a coverage of only conducting molecules; however, such a situation 
would result in low junction formation probability in the corresponding 
STM-BJ experiment. These shifts in conductance, due to collective surface 
effects, outweigh any contribution from the bulk solvent as it is not 
ordered: any ordering of the solvent, with an intrinsic dipole of the order 
of $\simeq1$ D in the presence of an electric field resulting from 
experimental bias of $\simeq1$ MV/cm are washed out by fluctuations at 
room temperature.  Moreover, the induced dipole upon binding is larger than 
the intrinsic dipole: 0 D vs. 7.6 D for BP and 2.5 D vs. 3.6 D for TCB. 
Only the molecules on the surface contribute to the junction potential: 
the surface templates their dipoles, leading to a coherent, collective 
effect. For experiments done at a lower temperature, higher bias, or 
with a solvent with a much higher intrinsic dipole, such bulk effects 
would be expected to begin to play an appreciable role, a subject for 
future study.

Finally, we support our theory with time-dependent conductance 
measurements of Au-BP-Au junctions, in the presence of TCB solvent 
(details in Supp. Mat.). We compare the percent change in conductance 
for $\Delta p_{0,s}$ = 3.6 D the dipole of TCB bound to gold to the 
percent change in conductance for BP on gold in a 0.01 mM solution of 
TCB over 5 hours. Over the course of the experiment the TCB evaporates 
and the concentration of BP in the vicinity of the junction increases, 
resulting in an increased conductance. As shown in Figure ~\ref{fig3}, 
we observe a variation of 20\% in conductance from our model 
($1.5\times 10^{-4} G_0-1.8\times 10^{-4}G_0$) and a variation of 37 
$\pm$ 12\% in the experiment 
($8.7\times 10^{-5}G_0 -1.11\times 10^{-4}G_0$). 
The overall absolute difference between experiment and 
theory can be attributed to the flat electrodes in the model and the 
lack of binding site variation, as we have assumed all adatom binding 
sites: by varying the binding site for a molecular coverage of 
0.8 mol/nm$^2$, we find a conductance of $1.8\times 10^{-4}G_0$ 
for adatom-adatom,  $1.2\times 10^{-4}G_0$ for adatom-trimer, and  
$9.6\times 10^{-5}G_0$ for trimer-trimer. 

In summary, we developed a quantitative understanding and general 
model of the effects of the solvent environment on the conductance 
of single-molecule junctions, and performed accompanying measurements. 
Our model predicts a significant shift in conductance for the specific 
case of Au-BP-Au junctions, with a magnitude and sign comparing very 
well with experiment. This solvent-induced electrostatic gating effect 
-- at its core -- is due to the solvent and conducting molecules 
bound to the surface at the vicinity of the junction, which changes 
the local electrostatic potential. We demonstrate that an electrostatic 
model approximating the junction and its surroundings by an array of 
point-polarizable dipoles quantitatively captures these effects and 
can be extended to incorporate the effect of nearby solvent molecules 
on the local potential and conductance, acting as a local gate. The 
magnitude of these reversible effects establishes liquid neutral 
solvents as a potential route to realize three-terminal device physics 
in single-molecular junctions. Our model and findings are general, 
and can be applied to arbitrary surface concentrations of solvent 
and conducting molecules, and is thus useful for predictive design 
of future multiterminal nanoscale transport devices. 

We thank Leeor Kronik for useful discussion and insight regarding the 
electrostatic model. This work was primarily performed at the Molecular 
Foundry and supported by the Division of Materials Sciences and 
Engineering (Theory FWP), under the auspices of the Office of Basic 
Energy Sciences of the U.S. Department of Energy under Contract No. 
DE-AC02-05CH11231. We also acknowledge support from AFOSR MURI 
FA9550-12-1-0002. Latha Venkataraman acknowledges support from the 
NSF DMR-1122594.

\bibliographystyle{apsrev4-1}
\bibliography{DipPaper}
\end{document}